\documentclass[aos,MSNbibl,nameyear,dvips]{arximspdf}
\usepackage{graphicx}

\doi{10.1214/12-AOS1058} %kopijuoti is PTS
\volume{41}
\issue{1}
\pubyear{2013}
\firstpage{196}
\lastpage{220}

\makeatletter
\newtheorem{proposition}{Proposition}
\newproclaim{definition}{Definition}
\newproclaim{property}{Property}
\makeatother

\begin{document}
\begin{frontmatter}

\title{On the definition of a confounder\thanksref{T1}}
\runtitle{Confounder definition}
\thankstext{T1}{Funded by the National Institutes of Health, USA.}

\begin{aug}
\author[A]{\fnms{Tyler J.} \snm{VanderWeele}\corref{}\ead[label=e1]{tvanderw@hsph.harvard.edu}\ead[label=u1,url]{http://www.hsph.harvard.edu/faculty/tyler-vanderweele/}}
\and
\author[B]{\fnms{Ilya} \snm{Shpitser}\ead[label=e2]{shpitse@hsph.harvard.edu}}
\runauthor{T.~J. VanderWeele and I. Shpitser}
\affiliation{Harvard University}
\address[A]{Departments
of Epidemiology\\
\quad and Biostatistics\\
Harvard School of Public Health\\
677 Huntington Avenue\\
Boston, Massachusetts 02115\\
USA\\
\printead{e1}\\
\printead{u1}}
\address[B]{Departments of Epidemiology\\
Harvard School of Public Health\\
677 Huntington Avenue\\
Boston, Massachusetts 02115\\
USA\\
\printead{e2}}
\end{aug}

% HISTORY:
\received{\smonth{12} \syear{2011}}
\revised{\smonth{9} \syear{2012}}

% ABSTRACT
%
\begin{abstract}
The causal inference literature has provided a clear formal definition of
confounding expressed in terms of counterfactual independence. The
literature has not, however, come to any consensus on a formal
definition of
a confounder, as it has given priority to the concept of confounding over
that of a confounder. We consider a number of candidate definitions
arising from various more informal statements made in the literature. We
consider the properties satisfied by each candidate definition, principally
focusing on (i) whether under the candidate definition control for all
``confounders'' suffices to control for ``confounding'' and (ii)
whether each
confounder in some context helps eliminate or reduce confounding bias.
Several of the candidate definitions do not have these two properties. Only
one candidate definition of those considered satisfies both properties. We
propose that a ``confounder'' be defined as a pre-exposure covariate
$C$ for
which there exists a set of other covariates $X$ such that effect of the
exposure on the outcome is unconfounded conditional on $(X,C)$ but such that
for no proper subset of $(X,C)$ is the effect of the exposure on the outcome
unconfounded given the subset. We also provide a conditional analogue
of the
above definition; and we propose a variable that helps reduce bias but not
eliminate bias be referred to as a ``surrogate confounder.'' These definitions
are closely related to those given by Robins and Morgenstern [\textit
{Comput. Math. Appl.} \textbf{14} (1987) 869--916].
The implications that hold among the various candidate definitions are
discussed.
\end{abstract}

% KEYWORDS
%


\end{frontmatter}

%s1 #&#
\section{Introduction}\label{sec1}
Statisticians and epidemiologists had traditionally conceived of a
confounder as a pre-exposure variable that was associated with exposure
and associated also with the outcome conditional on the exposure,
possibly conditional also on other covariates [\citet{Mie74}]. The
developments in causal inference over the past two decades have made
clear that this definition of a ``confounder'' is inadequate: there can
be pre-exposure variables associated with the exposure and the outcome,
the control of which introduces rather than eliminates bias [Greenland, Pearl and
Robins (\citeyear{GrePeaRob99}), \citet{GlyGre08}, \citet{Pea09}]. The
literature has moved away from formal language about ``confounders''
and instead places the conceptual emphasis on ``confounding.'' See
\citet{Mor11} for historical discussion of this point. The causal
inference literature has provided a formal definition of
``confounding'' in terms of dependence of counterfactual outcomes and
exposure, possibly conditional on covariates. The absence of
confounding (independence of the counterfactual outcomes and the
exposure) has been taken as the foundational assumption for drawing
causal inferences. Such absence of confounding is alternatively
referred to as ``ignorability'' or ``ignorable treatment assignment''
[\citet{Rub78}], ``exchangeability'' [\citet{GreRob86}], ``no
unmeasured confounding'' [\citet{Rob92}], ``selection on
observables'' [Barnow, Cain and
Goldberger (\citeyear{BarCaiGol80}), \citet{Imb04}] or
``exogeneity'' [\citet{Imb04}]. Today, at least within the formal
methodological literature on causality, language concerning
``confounders'' is generally used only informally, if at all. The
priority that has been given to ``confounding'' over ``confounders''
has arguably brought clarity and precision to the field. Nevertheless,
among practicing statisticians and epidemiologists, language concerning
both ``confounders'' and ``confounding'' is common. This raises the
question as to whether a formal definition of a ``confounder'' can also
be given within the counterfactual framework that coheres with how the
word seems to be used in practice.

In this paper we will consider various definitions of a confounder
proposed either formally or informally by a number of prominent
statisticians and epidemiologists. For each potential definition we
will consider the properties satisfied by the candidate definition.
Specifically, we state and prove a number of propositions showing
whether under each candidate definition (i) control for all
``confounders'' suffices to control for ``confounding'' and (ii)
whether each
confounder in some context helps eliminate or reduce confounding bias.
As we will see below, only
one candidate definition of those considered satisfies both properties.
We consider also the implications that hold between the various
definitions themselves.

%s2 #&#
\section{Notation and framework}\label{sec2}

We let $A$ denote an exposure, $Y$ the outcome, and we will use $C$,
$S$ and
$X$ to denote particular pre-exposure covariates or sets of covariates (that
may or may not be measured). As noted in the penultimate section of the
paper, the restriction to pre-exposure covariates could, in the context of
causal diagrams [Pearl (\citeyear{Pea95,Pea09})], be replaced to that of nondescendents
of exposure $A$. Within the counterfactual or potential outcomes
framework [\citet{Ney23}, \citet{Rub78}], we let $Y_{a}$ denote the
potential outcome for $Y$ if exposure $A$ were set, possibly contrary
to fact, to the value $a$. If the exposure is binary, the average
causal effect is given by $E(Y_{1})-E(Y_{0})$. Note that the potential
outcomes notation $Y_{a}$
presupposes that an individual's potential outcome does not depend on the
exposures of other individuals. This assumption is sometimes referred to
as SUTVA, the stable unit treatment value assumption [\citet{Rub90}] or
as a
no-interference assumption [\citet{Cox58}].

We use the notation $E\perp\!\!\!\perp F|G$ to denote that $E$ is independent
of $F$ conditional on $G$. For exposure $A$ and outcome $Y$, we say
there is
no confounding conditional on $S$ (or that the effect of $A$ on $Y$ is
unconfounded given $S$) if $Y_{a}\perp\!\!\!\perp A|S$. We will refer to any
such $S$ as a sufficient set or a sufficient adjustment set. If the
effect of $A$ on $Y$ is unconfounded given $S$, then the causal effect can
be consistently estimated by
$E(Y_{1})-E(Y_{0})= \sum_{s}\{E(Y|A=1,s)-E(Y|A=0,s)\}\operatorname{pr}(s)$ [\citet{RosRub83}].
We
will say that $S=(S_{1},\ldots,S_{n})$ constitutes a minimally sufficient
adjustment set if $Y_{a}\perp\!\!\!\perp A|S$ but there is no proper
subset $T$ of $S$ such that $Y_{a}\perp\!\!\!\perp A|T$, where ``proper
subset'' here
is understood as $T$ being a strict subset of the coordinates of
$S=(S_{1},\ldots,S_{n})$.

Some of the candidate definitions of a confounder
below define ``confounder'' in terms of ``confounding'' via reference
to ``sufficient adjustment sets'' or ``minimally sufficient adjustment
sets.'' Such definitions give conceptual
priority to ``confounding,'' as has generally been done in the causal
inference literature [\citet{GreRob86}, \citet{GreMor01}, Hern\'{a}n (\citeyear{Her08})]. Often after formal definitions of
``confounding'' are given, a ``confounder'' is defined as a derivative and
sometimes informal concept. For example, in papers by Greenland, Pearl and
Robins
(\citeyear{GrePeaRob99}) and \citet{GreMor01}, formal definitions are given
for ``confounding'' and then a ``confounder'' is simply described as a variable
that is in some sense ``responsible'' [Greenland, Robins and
Pearl (\citeyear{GreRobPea99}), page
33] for
confounding. Although priority arguably has and should be given to the
concept of ``confounding'' over ``confounder,'' applied researchers
will often use the word ``confounder'' to refer to a single variable
that is perhaps a member of a sufficient adjustment set but does not by
itself constitute a sufficient adjustment set and this raises the
question of whether this use of ``confounder'' can be given a coherent
definition within the counterfactual framework.

Most of the definitions and properties we discuss make reference only to
counterfactual outcomes. However, one of the definitions and several
propositions make reference to causal diagrams. We will thus restrict
attention in this paper to causal diagrams. We review concepts and
definitions for causal diagrams in the \hyperref[app]{Appendix}; the reader can also consult
Pearl (\citeyear{Pea95,Pea09}). For expository purposes we follow \citet{Pea95}, but
the results in the paper are equally applicable to all of the
alternative graphical causal models considered, for example, by \citet{RobRic10}. In short, following \citet{Pea95}, a causal
diagram is a very general data\vadjust{\goodbreak}
generating process corresponding to a set of nonparametric structural
equations where each variable $X_{i}$ is given by its nonparametric
structural equation $X_{i}=f_{i}(pa_{i},\varepsilon_{i})$, where
$pa_{i}$ are
the parents of $X_{i}$ on the graph and the $\varepsilon_{i}$ are mutually
independent such that the structural equations encode one-step ahead
counterfactual relationships among the variables with other
counterfactuals given by recursive substitution [Pearl (\citeyear{Pea95,Pea09})]. The
assumption of ``faithfulness'' is said to be satisfied if all of the
conditional independence relationships among the variables are implied by
the structure of the graph; see the \hyperref[app]{Appendix} for further details. A
backdoor path from $A$ to $Y$ is a path to $Y$ which begins with an edge
into $A$. \citet{Pea95} showed that if a set of pre-exposure covariates $S$
blocks all backdoor paths from $A$ to $Y$, then the effect of $A$ on
$Y$ is
unconfounded given~$S$.

The definitions given below will be stated formally in terms of
potential outcomes and causal
diagrams. It is assumed that there is an underlying causal diagram
which may
contain both measured and unmeasured variables; all variables
considered in
the definitions are variables on the diagram. Whether a variable satisfies
the criteria of a particular definition will be relative to the causal
diagram. In Section~\ref{sec6} we will consider settings with multiple causal
diagrams where one diagram may have variables absent on another.

%s3 #&#
\section{Candidate definitions for a confounder}\label{sec3}

Here we give a number of candidate definitions of a confounder
motivated by
statements made in the methodological literature. We will cite specific
statements from the methodologic literature; we do not necessarily believe
these statements were intended as formal definitions of a
``confounder'' by the
authors cited. We simply use these statements to motivate the candidate
definitions. As noted above, we believe statements about
``confounders,'' as
opposed to ``confounding,'' have generally been used only informally and
intuitively.

As already noted, the traditional conception of a confounder in
statistics and epidemiology
has been a variable associated with both the treatment and the outcome.
\citet{Mie74} notes that whether such associations hold will depend on
what other variables are controlled for in an analysis. This motivates our
first candidate definition for a confounder.

%de1 #&#
\begin{definition}\label{def1}A pre-exposure covariate $C$ is a confounder
for the effect of $A$ on $Y$ if there exists a set of pre-exposure
covariates $X$ such that $%
%TCIMACRO{\TeXButton{TeX field}{C \not\perp\!\!\!\perp A \mid X}}%
%BeginExpansion
C \not\perp\!\!\!\perp A \mid X%
%EndExpansion
$ and $%
%TCIMACRO{\TeXButton{TeX field}{C \not\perp\!\!\!\perp Y \mid(A,X)}}%
%BeginExpansion
C \not\perp\!\!\!\perp Y \mid(A,X).$
\end{definition}

Definition~\ref{def1} is essentially a generalization of the traditional
conceptualization of a confounder.

\citet{Pea95} showed that if a set of pre-exposure covariates $X$ blocks all
backdoor paths from $A$ to $Y$, then the effect of $A$ on $Y$ is unconfounded
given $X$. Hern\'{a}n (\citeyear{Her08}) accordingly speaks of a confounder as a
variable that ``can be used to block a backdoor path between exposure and
outcome'' (page 355). A similar definition of a confounder is given in
Greenland and Pearl [(\citeyear{GrePea07}), page 152] and in Glymour and Greenland
[(\citeyear{GlyGre08}), page
193]. This motivates a second candidate definition.

%de2 #&#
\begin{definition}\label{def2}A pre-exposure covariate $C$ is a confounder
for the effect of $A$ on $Y$ if it blocks a backdoor path from $A$ to $Y$.
\end{definition}

The second definition is perhaps one that would arise most naturally within
the context of causal diagrams; the definition itself of course presupposes
a framework of causal diagrams or variants thereof [Spirtes, Glymour and
Scheines
(\citeyear{SpiGlySch93}),
\citet{Daw02}].

\citet{Pea09} speaks of a confounder as ``a variable that is a member of every
sufficient [adjustment] set'' (page 195), that is, control for it must
be necessary.
Likewise, \citet{RobGre86} write, ``We will call a covariate a
confounder if estimators which are not adjusted for the covariate are
biased'' (page~393) and Hern\'{a}n (\citeyear{Her08}) speaks of a confounder as ``any
variable that is necessary to eliminate the bias in the analysis''
(page 357).
Note that a variable is a member of every sufficient adjustment set if and
only if it is a member of every minimal sufficient adjustment set. This
motivates our third candidate definition.

%de3 #&#
\begin{definition}\label{def3}A pre-exposure covariate $C$ is a confounder
for the effect of $A$ on $Y$ if it is a member of every minimally sufficient
adjustment set.
\end{definition}

Definition~\ref{def3} captures the notion that controlling for a confounder
might be
necessary to eliminate bias. The definition makes reference to ``every
minimally sufficient adjustment set;'' this will be relative to a particular
causal diagram, a point to which we will return below.

Kleinbaum, Kupper and
Morgenstern (\citeyear{KleKupMor82}), in a textbook on epidemiologic research, gave
as a
definition of a ``confounder'' a variable that is ``a member of a sufficient
confounder group'' where a sufficient confounder group is defined as ``a~minimal
set of one or more risk factors whose simultaneous control in the
analysis will correct for joint confounding in the estimation of the effect
of interest'' (page~276). Kleinbaum, Kupper and
Morgenstern (\citeyear{KleKupMor82}), however, define
``confounding'' in terms of association rather than counterfactual
independence. As a variant of the Kleinbaum, Kupper and
Morgenstern proposal, we could retain
the definition ``a member of a minimally sufficient adjustment set''
but use
the counterfactual definition of ``confounding.'' This motivates the fourth
candidate definition.

%
%de4 #&#
\begin{definition}\label{def4}A pre-exposure covariate $C$ is a confounder
for the effect of $A$ on $Y$ if it is a member of some minimally sufficient
adjustment set.
\end{definition}

Definition~\ref{def4} can be restated as follows: a pre-exposure covariate $C$
is a
confounder for the effect of $A$ on $Y$ if there exists a set of
pre-exposure covariates $X$ (possibly empty) such that $Y_{a}\perp\!\!\!
\perp A|(X,C)$ but
there is no proper subset $T$ of $(X,C)$ such that $Y_{a}\perp\!\!\!
\perp A|T$.
\citet{RobMor87} and \citet{Daw02} likewise conceive of a
confounder in terms of the presence or absence of confounding in such a
way that coincides with Definition~\ref{def4} when there is a single confounder;
when there are multiple sets that are sufficient or sets that are
sufficient but not minimally sufficient, it is not clear how the
definition of \citet{Daw02} generalizes; the definitions of \citet{RobMor87} can be adapted to coincide with Definition~\ref{def4}. Robins
and Morgenstern [(\citeyear{RobMor87}), Section~2H] say that $C$ is a confounder
conditional on $F$ if causal effects are computable given data on $C$
and $F$, but not on $F$ alone. In the framework of Robins and
Morgenstern, if one were to take as the (unconditional) definition of a
confounder that ``there exists some set $F$ such that $C$ is a
confounder conditional on $F$ [in the sense of \citet{RobMor87}, Section~2H],'' then this would coincide with Definition~\ref{def4}.

\citet{MieCoo81} conceive of a
confounder as any variable that is helpful in reducing bias. Hern\'{a}n
(\citeyear{Her08}) likewise speaks of a confounder as ``any variable that can be
used to
reduce [confounding] bias'' (page 355). Geng, Guo and Fung (\citeyear{GenGuoFun02}) use a similar
definition for confounding. As noted by other authors [\citet{GreMor01}, Hern\'{a}n (\citeyear{Her08})], whether a variable is helpful in
reducing bias will depend on what other variables are being conditioned on
in the analysis; a confounder should be helpful for reducing bias in some
context. This motivates our fifth definition.

%de5 #&#
\begin{definition}\label{def5}A pre-exposure covariate $C$ is a confounder
for the effect of $A$ on $Y$ if there exists a set of pre-exposure
covariates $X$ such that $| \sum_{x,c}\{E(Y|A=1,x,c)-E(Y|A=0,x,c)\}
\operatorname{pr}(x,c)-\{E(Y_{1})-E(Y_{0})\}|<\break| \sum_{x}\{E(Y|A=1,x)-E(Y|A=0,x)\}
\operatorname{pr}(x)-\{E(Y_{1})-E(Y_{0})\}|$.
\end{definition}

Definition~\ref{def5} captures the notion that controlling for $C$ along with $X$
results in lower bias in the estimate of the causal effect than controlling
for $X$ alone. A number of variants of Definition~\ref{def5} could also be
considered. Geng, Guo and Fung (\citeyear{GenGuoFun02}), for example, considered the analogous
definition for the effect of the exposure on the exposed rather than the
overall effect of the exposure on the population; one could likewise
consider the analogue of Definition~\ref{def5} for effects conditional on $X$
rather than standardized over $X$ or, alternatively, for different
measures of
effect, for example, risk ratios or odds ratios rather than causal
effects on the
difference scale. Definition~\ref{def5}, unlike other definitions, is inherently
scale-dependent. Thus, under Definition~\ref{def5}, a variable $C$ might be a
confounder for $Y$ but not for $\log(Y)$ or vice versa. This is an
important limitation of Definition~\ref{def5}.\vadjust{\goodbreak} Note, however, that some authors also
consider ``confounding'' to be scale-dependent [Greenland and Robins (\citeyear{GreRob86},
\citeyear{GreRob09}), \citet{GreMor01}] and use ``ignorability'' to
refer to
the notion of unconfoundedness in the distribution of counterfactuals as
given above.

Confounders have also
sometimes been defined in terms of empirical collapsibility [\citet{Mie76}, \citet{BreDay}], that is, if one obtains the same
estimate with or without adjustment for a variable, then it is not a
confounder. In the applied literature the approach is sometimes
encapsulated in the ``10 percent rule,'' that is, discard a covariate if
adjustment for it does not change an estimate by more than 10 percent.
It is well documented in the literature that collapsibility-based
definitions do not work for all
effect measures, such as the odds ratio or hazard ratios, for which
marginal and conditional may differ even in the absence of confounding
[Greenland, Robins and
Pearl (\citeyear{GreRobPea99})]. Such effect measures are sometimes referred
to as noncollapsible. However, for at least the risk difference scale
(or the risk ratio scale) a collapsibility-based definition of a
confounder could be entertained and for completeness we consider it
also here. Such a collapsibility-based definition could be formalized
as follows.

%de6 #&#
\begin{definition}\label{def6}A pre-exposure covariate $C$ is a confounder
for the effect of $A$ on $Y$ if there exists a set of pre-exposure
covariates $X$ such that $ \sum_{x,c}\{E(Y|A=1,x,c)-E(Y|A=0,x,c)\}
\operatorname{pr}(x,c)\neq\sum_{x}\{E(Y|A=1,x)-\break E(Y|A=0,x)\}\operatorname{pr}(x)$.
\end{definition}

Definition~\ref{def6}, like Definition~\ref{def5}, is scale-dependent.

Although not the focus of the present paper, in the \hyperref[app]{Appendix} we give some
further remarks on the possibility of empirical testing for each of
Definitions~\ref{def1}--\ref{def6} and for confounding and nonconfounding more generally.
However, for the most part, notions of confounding and confounders,
under these six definitions, are not empirically testable without
further experimental data or strong assumptions.

%s4 #&#
\section{Properties of a confounder}\label{sec4}

Language about ``confounders'' occurs of course not simply in methodologic
work but in substantive statistical and epidemiologic research. In the
design and analysis
of observational studies in the applied literature the task of
controlling for ``confounding'' is often construed as that of
collecting data
on and controlling for all ``confounders.'' In this section we propose that
when language about ``confounders'' is generally used in statistics and
epidemiology, two
things are implicitly presupposed: first, that if one were to control for
all ``confounders,'' then this would suffice to control for
``confounding'' and,
second, that control for a ``confounder'' will in some sense help to
reduce or
eliminate confounding bias. We would propose that if a formal\vadjust{\goodbreak}
definition is
to be given for a ``confounder,'' it should in some sense satisfy these two
properties. If it does not, it arguably does not cohere with what is
typically presupposed when language about ``confounders'' is used in
practice. We give a formalization of these two properties and
in the following section we will discuss which of these two properties are
satisfied by each of the candidate definitions of the previous section.

We could formalize the first property as follows.

%pr1 #&#
\begin{property}\label{prope1}If $S$ consists of the set of all confounders
for the effect of $A$ on~$Y$, then there is no confounding of the
effect of $%
A$ on $Y$ conditional on $S$, that is, $Y_{a}\perp\!\!\!\perp A|S$.
\end{property}

The definition makes reference to ``all confounders;'' to make
reference to
all such variables, the domain of the variables considered needs to be
specified. The domain here will be all pre-exposure variables on a
particular causal diagram that qualify as confounders according to whatever
definition is in view. See Section~\ref{sec6} for some extensions.

The second property is that control for a confounder should help either
reduce or eliminate bias. The reduction and the elimination of bias are not
equivalent and, thus, we will formally give two alternative properties,
\ref{prope2A} and
\ref{prope2B}.

\renewcommand{\theproperty}{\arabic{property}A}
%pr2 #&#
\begin{property}\label{prope2A}If $C$ is a confounder for the effect of $A$
on $Y$, then there exists a set of pre-exposure covariates $X$
(possibly empty) such that $%
Y_{a}\perp\!\!\!\perp A|(X,C)$ but $Y_{a}%
%TCIMACRO{\TeXButton{TeX field}{\not\perp\!\!\!\perp A \mid X}}%
%BeginExpansion
\not\perp\!\!\!\perp A \mid X%
%EndExpansion
$.
\end{property}

%pr3 #&#
\renewcommand{\theproperty}{\arabic{property}B}
\setcounter{property}{1}
\begin{property}\label{prope2B}If $C$ is a confounder for the effect of $A$
on $Y$, then there exists a set of pre-exposure covariates $X$
(possibly empty) such that $| \sum_{x,c}\{E(Y|A=1,x,c)-E(Y|A=0,x,c)\}
\operatorname{pr}(x,c)-\{E(Y_{1})-E(Y_{0})\}|<\break | \sum_{x}\{E(Y|A=1,x)-E(Y|A=0,x)\}
\operatorname{pr}(x)-\{E(Y_{1})-E(Y_{0})\}|$.
\end{property}

Property~\ref{prope2A} captures that notion that in some context, that is,
conditional on~$X$, the covariate $C$ helps eliminate bias. Property~\ref{prope2B}
captures the
notion that in some context, that is, conditional on $X$, the covariate $C$
helps reduce bias. Note that Property~\ref{prope2B}, like Definition~\ref{def5}, is inherently
scale-dependent and in this sense perhaps less fundamental than
Property~\ref{prope2A}.
For now we simply propose that for a candidate definition of a
confounder to
adequately capture the intuitive sense in which the word is used, it
should satisfy Property~\ref{prope1} and
should also satisfy either Property~\ref{prope2A} or~\ref{prope2B}. It would be peculiar if a
confounder were defined in a way that it did not satisfy these two
properties. In the next section we
consider whether each of the candidate definitions, Definitions~\ref{def1}--\ref{def6}, satisfy
Properties~\ref{prope1},~\ref{prope2A} and~\ref{prope2B}. Of course, one possible outcome of this exercise
is that none of the candidate definitions satisfy Property~\ref{prope1} and either Properties~\ref{prope2A}
or~\ref{prope2B} (or even that no candidate definition could). However, as we will see
in the next section, this turns out not to be the case.

%s5 #&#
\section{Properties of the candidate definitions}\label{sec5}

Definition~\ref{def1} was a generalization of the traditional epidemiologic
conception of a confounder as a variable associated with exposure and
outcome. For this definition we have the following result.

%pr1 #&#
\begin{proposition}\label{prop1}Under faithfulness, for every causal
diagram, Definition~\ref{def1} satisfies Property~\ref{prope1}. Definition~\ref{def1} does not satisfy
Properties~\ref{prope2A} or~\ref{prope2B}.
\end{proposition}

\begin{pf}We first show that Definition~\ref{def1} satisfies
Property~\ref{prope1} in faithful models.

Let $G^{\ast}=G_{\mathrm{Nd}(A)\cup \operatorname{An}(Y)}$ be the subgraph of $G$ that has only the nodes
in $\operatorname{Nd}(A)$ or $\operatorname{An}(Y)$; see the \hyperref[app]{Appendix}. Let
$\operatorname{Pa}^{\ast}$ be the subset of $\operatorname{Pa}(A)$ in $G^{\ast}$ such that every element
$P\in \operatorname{Pa}^{\ast}$ contains some path in $G^{\ast}$ to $Y$ not through $A$.
Since we consider faithful models, we can use d-connectedness to represent
dependence. First we note that every element in $\operatorname{Pa}^{\ast}$ satisfies
Definition~\ref{def1}. Indeed, any element of $\operatorname{Pa}(A)$ is dependent on $A$ conditioned
on any set. For any member of $\operatorname{Pa}^{\ast}$, we fix some path $\pi$ to $Y$
(not through $A$). We are now free to pick any set $X$ to make this path
d-connected (e.g., we can pick the smallest $X$ that opens all
colliders in $\pi$). This set $X$ satisfies Definition~\ref{def1} for $\operatorname{Pa}^{\ast}$
with respect to $A$ and $Y$. Thus, the set of all nodes in $\operatorname{Nd}(A)$
satisfying Definition~\ref{def1} will include $\operatorname{Pa}^{\ast}$. Next, we show that any
superset of $\operatorname{Pa}^{\ast}$ in $\operatorname{Nd}(A)$ will be a valid adjustment set for
$(A,Y) $. Assume this is not the case for a particular $S$, and fix a
backdoor path from $A$ to $Y$ which is open given $S$. Then the first node
on this path after $A$ must be in $\operatorname{Pa}^{\ast}$. But this means the path is
blocked by $S$. Our conclusion follows.

%f1 #&#
\begin{figure}

\includegraphics{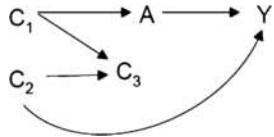}

\caption{Definition \protect\ref{def1} does not satisfy Property \textup{\protect\ref{prope2A}} or \textup{\protect\ref{prope2B}}.}\label{fig1}
\end{figure}

We now show Definition~\ref{def1} does not satisfy Properties~\ref{prope2A} or~\ref{prope2B}. Consider the
causal diagram in Figure~\ref{fig1}. The variable $C_{3}$ is unconditionally
associated with $A$ and $Y$; the variables $C_{1}$ and $C_{2}$ are each
associated with $A$ and $Y$ conditional on $C_{3}$. Thus, under
Definition~\ref{def1},
all three would qualify as ``confounders.'' However, there is no set of
pre-exposure
covariates $X$ on the graph such that control for $C_{3}$ helps
eliminate or reduce bias. To see this, note that if $X$ includes
$C_{1}$ or $C_{2}$, then the effect estimate is unbiased irrespective
of whether adjustment is made for $C_{3}$. If $X$ includes neither
$C_{1}$ nor $C_{2}$, then the estimand without adjustment for $C_{3}$
is unbiased whereas the estimand adjusted for $C_{3}$ is not.
Therefore, Definition~\ref{def1} does not satisfy Properties~\ref{prope2A} or~\ref{prope2B}. This
completes the proof.
\end{pf}

Intuitively, Definition~\ref{def1} does not satisfy Properties~\ref{prope2A} or~\ref{prope2B} because
in the causal diagram in Figure~\ref{fig1}, the variable $C_{3}$ is
unconditionally associated with $A$ and $Y$ and thus would be a
confounder under Definition~\ref{def1}, but control for it will only either not
affect bias (if control is not made for $C_{1}$ and $C_{2}$) or
increase bias (if control is not made for $C_{1}$ and $C_{2}$). The
causal structure in Figure~\ref{fig1} and the bias resulting from controlling
for $C_{3}$ is sometimes referred to in the literature as
``M-bias'' or ``collider-stratification'' [\citet{Gre03}, Hern\'{a}n
et al.
(\citeyear{Heretal02}), Hern\'{a}n (\citeyear{Her08})]. We note that if faithfulness is violated,
Definition~\ref{def1} does not satisfy
Property~\ref{prope1} either [\citet{Pea09}].

Under Definition~\ref{def2}, a confounder was defined as a pre-exposure covariate
that blocks a backdoor path from $A$ to $Y$.

%pr2 #&#
\begin{proposition}\label{prop2}For every causal diagram, Definition~\ref{def2}
satisfies Property~\ref{prope1}. Definition~\ref{def2} does not satisfy Properties~\ref{prope2A} or~\ref{prope2B}.
\end{proposition}

\begin{pf}If $S$ consists of the set of all
confounders under Definition~\ref{def2}, then this set $S$ will include all
pre-exposure covariates that block a backdoor path from $A$ to $Y$. From
this it follows that $S$ blocks all backdoor paths from $A$ to $Y$ and by
Pearl's backdoor path theorem, the effect of $A$ on $Y$ is unconfounded
given $S$. Thus, Definition~\ref{def2} satisfies Property~\ref{prope1}.

%f2 #&#
\begin{figure}

\includegraphics{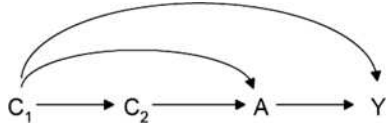}

\caption{Definition \protect\ref{def2} does not satisfy Property \textup{\protect\ref{prope2A}} or \textup{\protect\ref{prope2B}}.}\label{fig2}
\end{figure}

We now show that it does not satisfy Properties~\ref{prope2A} and~\ref{prope2B}. Consider the
causal diagram in Figure~\ref{fig2}. Under Definition~\ref{def2} both $C_{1}$ and $C_{2}$
block a backdoor path from $A$ to $Y$ and thus would qualify as
confounders. However, for $C_{2}$ there is no set of pre-exposure
covariates $X$ on the graph such that control for $C_{2}$ helps
eliminate since if $X=C_{1}$, there is no bias without controlling for
$C_{2}$; if $X=\varnothing$, there is bias even with controlling for
$C_{2}$. Thus,
Definition~\ref{def2} does not satisfy Property~\ref{prope2A}. We now show that it does not
satisfy Property~\ref{prope2B}. Suppose Figure~\ref{fig2} is a causal diagram for
$(C_{1},C_{2},A,Y)$ where all variables are binary and suppose that
$P(C_{1}=1)=1/2$, $P(C_{2}=1|c_{1})=1/5+3c_{1}/5$,
$P(A=1|c_{1},c_{2})=1/10+3c_{1}/5+c_{2}/10$,
$P(Y=1|a,c_{1},c_{2})=1/2+(1/2)(a-1/2)c_{1}$. One can then verify that
$E(Y_{1})-E(Y_{0})=\sum_{c_{1},c_{2}}\{
E(Y|A=1,c_{1},c_{2})-E(Y|A=0,c_{1},c_{2})\}\operatorname{pr}(c_{1},c_{2})=0.25
=\sum_{c_{1}}\{E(Y|A=1,c_{1})-E(Y|A=0,c_{1})\}\operatorname{pr}(c_{1})$, that
$E(Y|A=1)-E(Y|A=0)=0.266$ and that $\sum_{c_{2}}
\{E(Y|A=1,c_{2})-E(Y|A=0,c_{2})\}\operatorname{pr}(c_{2})=0.269$. Under Definition~\ref{def2},
$C_{2} $ would be considered a confounder since $C_{2}$ blocks the backdoor
path $A\leftarrow C_{2}\leftarrow C_{1}\rightarrow Y$. However, there
is no
set $X$ of pre-exposure covariates such that $|\sum_{x,c_{2}}\{
E(Y|A=1,x,c_{2})-E(Y|A=0,x,\break c_{2})\}\operatorname{pr}(x,c_{2})-\{E(Y_{1})-E(Y_{0})\}
|<|\sum_{x}\{E(Y|A=1,x)-E(Y|A=\break0,x)\}\operatorname{pr}(x)-\{E(Y_{1})-E(Y_{0})\}|$.
This is because if $X$ is taken as $C_{1}$, then the expressions on both
sides of the inequality are equal to $0$ (controlling for $C_{2}$ in
addition to $C_{1}$ does not reduce bias); if $X$ is taken as the empty set,
we have $|\sum_{c_{2}}\{E(Y|A=1,c_{2})-E(Y|A=0,c_{2})\}\operatorname{pr}(c_{2})-\{
E(Y_{1})-E(Y_{0})\}|=|0.269-0.250|=0.019>0.016=|0.266-0.250|=|\{
E(Y|A=1)-E(Y|A=0)\}-\{E(Y_{1})-E(Y_{0})\}|$ and again controlling for
$C_{2}$ does not reduce (but rather increases) bias. Definition~\ref{def2} thus does
not satisfy Property~\ref{prope2B}. This completes the proof.
\end{pf}

If we consider the causal diagram in Figure~\ref{fig2}, then under Definition~\ref{def2}
both $C_{1}$
and $C_{2}$ block a backdoor path from $A$ to $Y$ and thus would
qualify as
confounders. However, for $C_{2}$ there is no set of pre-exposure
covariates $X$ on the graph such that control for $C_{2}$ helps eliminate
bias (Property~\ref{prope2A}) since if $X=C_{1}$, there is no bias without controlling
for $
C_{2}$; if $X=\varnothing$, there is bias even with controlling for
$C_{2}$. Likewise, examples can be constructed as in the proof above in which
control for $C_{2}$ will only increase bias, that is, control for
$C_{2}$ does
not help reduce bias (Property~\ref{prope2B}).

%f3 #&#
\begin{figure}

\includegraphics{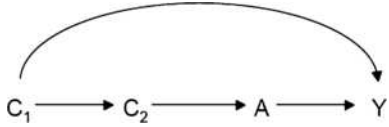}

\caption{Definition \protect\ref{def3} does not satisfy Property \protect\ref{prope1}.}\label{fig3}
\end{figure}

Under Definition~\ref{def3}, a confounder was defined as a member of every minimally
sufficient adjustment set.

%pr3 #&#
\begin{proposition}\label{prop3}Definition \protect\ref{def3} does not satisfy Property~\ref{prope1}.
Definition \protect\ref{def3} satisfies Property~\ref{prope2A}.
\end{proposition}

\begin{pf}Consider the causal diagram in Figure~\ref{fig3}.
Here, either $C_{1}$ or $C_{2}$ would constitute minimally sufficient
adjustment sets and thus neither are a member of every minimally sufficient
adjustment set and under Definition~\ref{def3}, neither would be confounders. If we
control for nothing, there is still confounding for the effect of $A$
on $Y$
and, thus, for Figure~\ref{fig3}, controlling for all confounders under
Definition~\ref{def3}
would not suffice to control for confounding. Thus, Definition~\ref{def3} does not
satisfy Property~\ref{prope1}. If $C$ is a member of every minimally sufficient
adjustment set, then it is a member of a minimally sufficient
adjustment set
and from this it trivially follows that it satisfies the requirements in
Property~\ref{prope2A}. This completes the proof.
\end{pf}

A variable $C$ that is a confounder under Definition~\ref{def3} will in general
satisfy Property~\ref{prope2B} as well but may not always because there are cases in
which there is confounding in the distribution\vadjust{\goodbreak} of counterfactual outcomes
conditional on $C$ and so that $C$ is a confounder under Definition~\ref{def3} but
with the average causal effect on the additive scale not confounded
[Greenland, Robins and
Pearl (\citeyear{GreRobPea99})]. Intuitively, to see that Definition~\ref{def3} does
not satisfy
Property~\ref{prope1}, consider the causal diagram in Figure~\ref{fig3}. Here, either
$C_{1}$ or $C_{2}$
would constitute minimally sufficient adjustment sets and thus neither
are a
member of every minimally sufficient adjustment set. Under Definition~\ref{def3},
there would thus be no confounders for the effect of $A$ on $Y$; clearly,
however, if we control for nothing, there is still confounding for the effect
of $A$ on $Y$.

Under Definition~\ref{def4}, a confounder was defined as a member of some minimally
sufficient adjustment set.

%pr4 #&#
\begin{proposition}\label{prop4}For every causal diagram, Definition~\ref{def4}
satisfies Property~\ref{prope1}. Definition~\ref{def4} satisfies Property~\ref{prope2A}.
\end{proposition}

\begin{pf}We will show that Definition~\ref{def4} satisfies
Property~\ref{prope1}. We first claim that any minimally sufficient adjustment set
for $(A,Y)$ must lie in $G_{\mathrm{An}(A)\cup \mathrm{An}(Y)}$,
the subgraph of $G$ that has only the nodes in $\operatorname{Nd}(A)$ or $\operatorname{An}(Y)$; see the \hyperref[app]{Appendix}. Assume this is not true,
and pick
some minimally sufficient set $S$ with elements outside $\operatorname{An}(A)\cup \operatorname{An}(Y)$.
This means $S\cap(\operatorname{An}(A)\cup \operatorname{An}(Y))$ is not sufficient. Note that any
ancestor of a node in the set $\operatorname{An}(A)\cup \operatorname{An}(Y)$ will also be in
$\operatorname{An}(A)\cup
\operatorname{An}(Y)$. From this it follows that any backdoor path from $A$ to $Y$ which
has a node outside $\operatorname{An}(A)\cup \operatorname{An}(Y)$ will require a collider to get back
into $\operatorname{An}(A)\cup \operatorname{An}(Y)$. However, those colliders must be open by
elements in
$S$. We have a contradiction. We have shown that any minimally sufficient
adjustment set must be a subset of $\operatorname{An}(A)\cup \operatorname{An}(Y)$ and, thus, any variable
that is a confounder under Definition~\ref{def4} must be in $\operatorname{An}(A)\cup \operatorname{An}(Y)$.

Next we note that $\operatorname{Pa}(A)$ is a sufficient adjustment set for $(A,Y)$.
Pick a
minimal subset $\operatorname{Pa}^{+}$ of $\operatorname{Pa}(A)$ that is sufficient. Our claim is that
every element $P$ in $\operatorname{Pa}(A)\setminus \operatorname{Pa}^{+}$ is such that $P$ is not
connected to $Y$ in the graph $(G_{\mathrm{An}(A)\cup \mathrm{An}(Y)})_{\overline{a}}$ except
by paths that are blocked conditional on $\operatorname{Pa}^{+}$. Assume this is not true,
and fix a path $\omega$ from $P$ to $Y$ that is not blocked by
$\operatorname{Pa}^{+}$ in $(G_{\mathrm{An}(A)\cup \mathrm{An}(Y)})_{\overline{a}}$. If this path has no
colliders, then
appending $\omega$ with the edge $P\rightarrow A$ produces a backdoor path
from $A$ to $Y$ not blocked by $\operatorname{Pa}^{+}$, contradicting the earlier claim
that $\operatorname{Pa}^{+}$ is a valid adjustment set.

If $\omega$ only contains colliders ancestral of $\operatorname{Pa}^{+}$, then either
$\omega$ has a noncollider triple blocked by $\operatorname{Pa}^{+}$ (in which case we are
done with that path) or $\omega$ appended with $P\rightarrow A$
produces a
backdoor path open conditional on $\operatorname{Pa}^{+}$, which is a contradiction.
If $\omega$ contains collider triples ancestral of $\operatorname{Pa}(A)\setminus
\operatorname{Pa}^{+}$ (but
not ancestral of $\operatorname{Pa}^{+}$), let $W$ be the central node of the last such
collider triple on the path from $P$ to $Y$. Let $P^{\prime}$ be a member
of $\operatorname{Pa}(A)\setminus \operatorname{Pa}^{+}$ of which $W$ is an ancestor. Consider
instead of $\omega$ a new path: $A\leftarrow P^{\prime}\leftarrow
\cdots\leftarrow W$\vadjust{\goodbreak}
appended with the subpath of $\omega$ that begins with the node on
$\omega$
after $W$ and ends with $Y$. This path either has a noncollider triple
blocked by $\operatorname{Pa}^{+}$ (in which case so does $\omega$ and we are done
with $\omega$) or it is open conditional on $\operatorname{Pa}^{+}$, in which case we
have a
contradiction, or it contains collider triples ancestral of $Y$ not
through $\operatorname{Pa}(A)$. In the last case, let $Z$ be the central node of the
first such
collider triple on the currently considered path from $A$ to $Y$. Consider
instead a new path which appends a subpath of the currently considered path
extending from $A$ to $Z$, and the segment $Z\rightarrow\cdots\rightarrow Y$.
This path has no blocked colliders by construction, and thus must either
have a noncollider triple blocked by $\operatorname{Pa}^{+}$ (in which case so does
$\omega$ and we are done with $\omega$) or it is open conditional on
$\operatorname{Pa}^{+}$, in which case we have a contradiction.

Our final claim is that any superset $S$ of $\operatorname{Pa}^{+}$ in $\operatorname{Nd}(A)\cap
(\operatorname{An}(A)\cup \operatorname{An}(Y))$ is a valid adjustment set for $(A,Y)$. Assume this were
not so and fix an open backdoor path $\rho$ from $A$ to $Y$ given $S$. The
first node on $\rho$ after $A$ must lie either in $\operatorname{Pa}^{+}$ or in
$\operatorname{Pa}(A)\setminus \operatorname{Pa}^{+}$. In the first case, the path is blocked. In the
second case, we have shown above that every path from $\operatorname{Pa}(A)\setminus \operatorname{Pa}^{+}$
to $Y$ in $(G_{\mathrm{An}(A)\cup \mathrm{An}(Y)})_{\overline{a}}$ is blocked by $\operatorname{Pa}^{+}$ and,
thus, the path must be blocked in the second case as well. There thus cannot
be an open backdoor path from $A$ to $Y$ given $S$ and we have a
contradiction. We have that $\operatorname{Pa}^{+}$ is a sufficient adjustment set; any
variable that is a confounder under Definition~\ref{def4} will be a member of
$\operatorname{Nd}(A)\cap(\operatorname{An}(A)\cup \operatorname{An}(Y))$ and, thus, we have that the set of
variables that
are confounders under Definition~\ref{def4} will be a sufficient adjustment set.
Definition~\ref{def4} thus satisfies Property~\ref{prope1}. Definition~\ref{def4} satisfies Property~\ref{prope2A} trivially. This completes the proof.
\end{pf}

A variable that is a confounder under Definition~\ref{def4} will in general satisfy
Property~\ref{prope2B} as well but may not always because, as before, there may be
confounding in distribution without the average causal effect on the
additive scale being confounded. Definition~\ref{def4} thus satisfies Property~\ref{prope2A},
generally Property~\ref{prope2B}, and, as shown in the proof above, also satisfies Property~\ref{prope1} for all causal diagrams. That Definition~\ref{def4} satisfies Property~\ref{prope1} can
be restated
as the proposition that the union of all minimally sufficient
adjustment sets is
itself a sufficient adjustment set. Definition~\ref{def4} thus satisfies the properties
which arguably ought to be required for a reasonable definition of a
``confounder.''

Under Definition~\ref{def5}, a confounder was essentially defined as a pre-exposure
covariate, the control for which helped reduce bias.

%pr5 #&#
\begin{proposition}\label{prop5}Definition~\ref{def5} does not satisfy Property~\ref{prope1}.
Definition~\ref{def5} satisfies Property~\ref{prope2B} but not~\ref{prope2A}.
\end{proposition}

\begin{figure}

\includegraphics{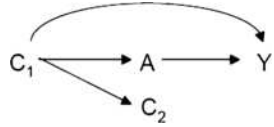}

\caption{Definition~\protect\ref{def5} does not satisfy Property \textup{\protect\ref{prope2A}}.}\label{fig4}
\end{figure}

\begin{pf}Suppose that $Y_{a}\perp\!\!\!\perp A|C$,
that $(C,A,Y)$ are all binary and that $P(C=1)=1/2$,
$P(A=1|c)=1/4+c/2$, $P(Y=1|a,c)=4/10-4c/10-3a/10+8ac/10$.\vadjust{\goodbreak} One can then
verify that $E(Y_{1})=\sum_{c}E(Y|A=1,c)\operatorname{pr}(c)=3/10$, $E(Y|A=1)=4/10$,
$E(Y_{0})=\sum_{c}E(Y|A=0,c)\operatorname{pr}(c)=2/10$, $E(Y|A=0)=3/10$. Thus,
$|\sum_{c}\{E(Y|A=1,c)-E(Y|A=0,c)\}\operatorname{pr}(c)-\break \{E(Y_{1})-E(Y_{0})\}|=0=|\{
E(Y|A=1)-E(Y|A=0)-\{E(Y_{1})-E(Y_{0})\}|$ and so under Definition~\ref{def5}, $C$
would not be a confounder. The set of variables defined as confounders under
Definition~\ref{def5} would thus be empty. However, it is not the case that
adjustment for the empty set suffices to control for confounding since, for
example, $E(Y_{1})=3/10\neq4/10=E(Y|A=1)$. Thus, Definition~\ref{def5} does not
satisfy Property~\ref{prope1}. We now show that Definition~\ref{def5} does not satisfy Property~\ref{prope2A}. Consider the causal diagram in Figure~\ref{fig4}. Although control for $C_{2}$
might reduce bias compared to an unadjusted estimate and thus satisfy
Definition~\ref{def5} with $X=\varnothing$, there is no $X$ such that the
effect of $A$ on $Y$ is unconfounded conditional on $(X,C_{2})$ but not
on $X$ alone.
Thus, Definition~\ref{def5} does not satisfy Property~\ref{prope2A}. Definition~\ref{def5} satisfies
Property~\ref{prope2B} trivially. This completes the proof.
\end{pf}

Definition~\ref{def5} does not satisfy Property~\ref{prope1} because an unadjusted estimate of
the causal risk difference may be correct, even in the presence of
confounding, because the bias due to confounding for $E(Y_{1})$ may cancel
that for $E(Y_{0})$; said another way, there may be confounding in the
distribution of counterfactual outcomes without their being confounding
in a
particular measure. That
Definition~\ref{def5} satisfies Property~\ref{prope2B} is essentially embedded in
Definition~\ref{def5}
itself. Intuitively, to see that Definition~\ref{def5} does not satisfy Property~\ref{prope2A}, consider the
causal diagram in Figure~\ref{fig4}. Although control for $C_{2}$
might reduce bias compared to an unadjusted estimate and thus satisfy
Definition~\ref{def5} with $X=\varnothing$, there would be no $X$ such that the
effect of $A$ on $Y$ is unconfounded conditional on $(X,C_{2})$ but not
on $X $ alone.

%f4 #&#

Under Definition~\ref{def6}, a confounder was defined as a pre-exposure
covariate, the control for which in some context changed the effect estimate.

%pr6 #&#
\begin{proposition}\label{prop6}Definition~\ref{def6} does not satisfy Property~\ref{prope1}. Definition~\ref{def6} does not satisfy Properties~\ref{prope2A} or~\ref{prope2B}.
\end{proposition}

\begin{pf}In the first example in the proof of Proposition~\ref{prop5}, the
set of confounders under Definition~\ref{def6} would be empty because with $X$ empty we have $ \sum_{x,c}\{E(Y|A=1,x,c)-E(Y|A=0,x,c)\}\operatorname{pr}(x,c)=0= \sum_{x}\{
E(Y|A=1,x)-E(Y|A=0,x)\}\operatorname{pr}(x)$. However,\vadjust{\goodbreak} the effect of $A$ on $Y$ is not
unconfounded conditional on the empty set. Thus, Definition~\ref{def6} does not
satisfy Property~\ref{prope1}.

We now show Definition~\ref{def6} does not satisfy Properties~\ref{prope2A} or~\ref{prope2B}. Consider the
causal diagram in Figure~\ref{fig1}. If we let $X$ denote the empty set, then
$C_{3}$ will satisfy Definition~\ref{def6} and so would be a confounder under
Definition~\ref{def6}. However, if we consider Properties~\ref{prope2A} and~\ref{prope2B}, there is no
set of pre-exposure
covariates $X$ on the graph such that control for $C_{3}$ helps
eliminate or reduce bias. To see this, note that if $X$ includes
$C_{1}$ or $C_{2}$, then the effect estimate is unbiased irrespective
of whether adjustment is made for $C_{3}$. If $X$ includes neither
$C_{1}$ nor $C_{2}$, then the estimand without adjustment for $C_{3}$
is unbiased whereas the estimand adjusted for $C_{3}$ is not.
Therefore, Definition~\ref{def1} does not satisfy Properties~\ref{prope2A} and~\ref{prope2B}. This
completes the proof.
\end{pf}

As with Definition~\ref{def5}, Definition~\ref{def6} does not satisfy Property~\ref{prope1} because
of the possibility of cancellations: there may be confounding in the
distribution of counterfactual outcomes without their being confounding
in a
particular measure. Definition~\ref{def6} also fails to satisfy Properties~\ref{prope2A} or~\ref{prope2B}. It
fails because of the possibility of ``M-bias'' or
``collider-stratification'' structures as in Figure~\ref{fig1}
[\citet{Gre03},
Hern\'{a}n et al. (\citeyear{Heretal02})]. Controlling for a variable such as
$C_{3}$ may change the estimate, but it may be that it is the estimate
without control for that variable (e.g., $C_{3}$ in Figure~\ref{fig1}) that is
unbiased. Also, as noted above, the collapsibility-based definitions
fail for odds ratio and hazard ratio measures for others reasons,
namely, because marginal and conditional measures are not comparable
even in the absence of confounding. See Greenland, Robins and
Pearl (\citeyear{GreRobPea99}), \citet{Genetal01} and
\citet{GenLi02} for further discussion of the
relationship between, and
general nonequivalence of, confounding and collapsibility.

Candidate definitions for a confounder might thus include Definition~\ref{def4}
and, if the issue of scale dependence is set aside, Definition~\ref{def5}. Note,
however, that a variable that satisfies Definition~\ref{def5} but not Definition~\ref{def4} will never help
to eliminate confounding bias, only to reduce such bias. Such a variable
reduces bias essentially by serving as a proxy for a variable that does
satisfy Definition~\ref{def4}. We therefore propose that a confounder be defined as
in Definition~\ref{def4}, ``a pre-exposure covariate that is a member of some
minimally sufficient adjustment set'' and that any variable that satisfies
Definition~\ref{def5} but not Definition~\ref{def4} be referred to as a ``surrogate
confounder.'' The terminology of a ``surrogate confounder'' or ``proxy
confounder'' appears elsewhere [\citet{GreMor01}, Hern\'{a}n
(\citeyear{Her08})]; here we have provided a formal criterion for such a ``surrogate
confounder.'' See \citet{GrePea11} and \citet{OgbVan12} for properties of such surrogate confounders.

Interestingly, Definition~\ref{def4} is closely related to definitions
concerning confounders proposed by Robins and Morgernstern (\citeyear{RobMor87}),
though their definitions were not universally adopted by the
epidemiologic community over the ensuing 25 years.
Robins and Morgenstern (\citeyear{RobMor87}) were not principally concerned with how the word
``confounder''
is employed in practice when used in an unqualified sense, but rather with whether
a particular variable would still, in some sense, be a confounder if data were also available on other variables.
As noted above,
Robins and Morgenstern [(\citeyear{RobMor87}), Section~2H] say that $C$ is a confounder
conditional on $F$ if causal effects are computable given data on $C$
and $F$, but not on $F$ alone. In the framework of Robins and
Morgenstern, if one were to take as the (unconditional) definition of a
confounder that ``there exists some set $F$ such that $C$ is a
confounder conditional on $F$ [in the sense of \citet{RobMor87}, Section~2H],'' then this would coincide with Definition~\ref{def4}. Note that
Robins and Morgenstern, in their definitions, in some sense go further
than Definition~\ref{def4} in having the investigator explicitly specify the
other variables $F$ for which control might be made. This would indeed
be useful in practice, though current use of language has not generally
adopted this convention. %In any case, the definition above does satisfy
%the properties typically presupposed by investigators when the word
%``confounder'' is generally used in practice, namely, Properties~\ref{prope1} and~\ref{prope2A}.
It might in the future be helpful to distinguish between the unqualified use of the word
``\textit{confounder}'' as defined in Definition~\ref{def4}, and ``\textit{confounder in the context of} having data also on
$F$''
as in Robins and Morgenstern (\citeyear{RobMor87}). The former is arguably how the word ``confounder'' is
often used in practice; the latter would be useful in making decisions about data collection and confounder control.

%s6 #&#
\section{Some extensions, implications and further results}\label{sec6}

In the discussion above we have considered whether a covariate is a
``confounder'' in an unconditional sense. However, we might also speak about
whether a variable $C$ is a confounder for the effect of $A$ on $Y$
conditional on some set of covariates $L$ which an investigator is
going to
condition on irrespective of whether control is made for~$C$.
Definition~\ref{def4}
above, the definition for an ``unconditional confounder'' could be restated
as follows: a pre-exposure covariate $C$ is a confounder for the effect
of $A$ on $Y$
if there exists a set of pre-exposure covariates $X$ such that
$Y_{a}\perp
\!\!\!\perp A|(X,C)$ but there is no proper subset $T$ of $(X,C)$ such
that $Y_{a}\perp\!\!\!\perp A|T$. The conditional analogue would then be as
follows: we say that a pre-exposure covariate $C$ is a confounder for the
effect of $A$ on $Y$ conditional on $L$ if there exists a set of
pre-exposure covariates $X$ such that $Y_{a}\perp\!\!\!\perp A|(X,L,C)$ but
there is no proper subset $T$ of $(X,C)$ such that $Y_{a}\perp\!\!\!\perp
A|(T,L)$. Consider again the causal diagram in Figure~\ref{fig3}. Here, $C_{2}$ would
be a confounder under Definition~\ref{def4}. However, $C_{2}$ is not a confounder
for the effect of $A$ on $Y$ conditional on $L=C_{1}$. Consider once more
the causal diagram in Figure~\ref{fig1}. Here, neither $C_{1}$ nor $C_{2}$ would be
a confounder under Definition~\ref{def4}. However, conditional on $L=C_{3}$,
both $C_{1}$ and $C_{2}$ would be confounders.

An analogue of Definition~\ref{def4} could also be given for a particular causal
parameter of interest rather than for the condition of nonconfounding
in distribution $Y_{a}\perp\!\!\!\perp A|S$.\vadjust{\goodbreak} For example, $C$ could be
defined to be a confounder for a particular causal parameter (e.g., the
causal risk difference or causal risk ratio) if there exists a set of
pre-exposure covariates $X$ such the parameter is identified by
adjusting for $(X,C)$ and if for no proper subset, $T$ of $(X,C)$ is
the parameter identified by adjusting for $T$ [cf. \citet{RobMor87}].
However, when we restrict attention to particular
parameters we reintroduce some of the complications with cancellations
that were noted above. For example, due to cancellations, a variable
$C$ may be a confounder for the causal risk difference but not for the
causal risk ratio [cf. \citet{Van12}].

We have restricted our attention in this paper thus far to pre-exposure
covariates as potential confounders. We have done so in order to
correspond as closely as possible to the discussion in the
epidemiologic and
potential outcomes literatures. However, within the context of causal
diagrams, a somewhat broader range of variables could be considered as
``confounders'' in that all of the discussion above is applicable if we
consider all nondescendents of $A$ as potential confounders rather than
simply considering pre-exposure covariates.

Throughout the paper we have given all definitions with respect to a
particular underlying causal diagram. However, for a given exposure $A$ and
a given outcome~$Y$, there will be multiple causal diagrams that correctly
represent the causal structure relating these variables to one another and
to covariates. One diagram may be an elaboration of another and contain
variables that the other does not. It is straightforward to verify that
if a
variable $C$ is classified as a confounder under Definitions~\ref{def1},~\ref{def2},~\ref{def4},~\ref{def5}
or~\ref{def6},
then $C$ will also be a confounder under each of those definitions respectively on any expanded
causal diagram with additional variables. In the case of Definition~\ref{def1}, this
is because associations that hold conditional on covariates $X$ for one
diagram will clearly also hold for the other. In the case of Definition~\ref{def2},
if $C$ blocks a backdoor path on one causal diagram, it will block a
backdoor path on any larger diagram that also correctly describes the causal
structure. In the case of Definition~\ref{def4}, if there is some minimally
sufficient adjustment set $S$ of which $C$ is a member, then that set will
also be minimally sufficient on any larger diagram that also correctly
describes the causal structure. In the case of Definitions~\ref{def5} and~\ref{def6}, if the
inequalities in these definitions hold for some covariate set $X$ for one
diagram, they will clearly also hold for the other. Only Definition~\ref{def3}
does not
share this property. To see this, consider Figure~\ref{fig3}; if in Figure~\ref{fig3}, we
collapsed over $C_{2}$ so that the causal diagram involved only
$C_{1}$, $A$ and $Y$, then $C_{1}$ would be a member of every minimally
sufficient
adjustment set for this diagram and thus a confounder under Definition~\ref{def3}.
However, as we saw above, $C_{1}$ is not a confounder under Definition~\ref{def3} for
Figure~\ref{fig3} itself which includes the extra variable $C_{2}$. This failure
is a
serious problem with Definition~\ref{def3}, but, as we also saw above,
Definition~\ref{def3}
suffers from other limitations as well.

Several fairly trivial implications follow from Definition~\ref{def4} and may be
worth noting for the sake of completeness. First, if a causal diagram
had a
variable $C$ with an arrow to $\log(C)$ (or vice versa) and if $C$
were a
member of a minimally sufficient adjustment set, then, under Definition~\ref{def4},
both $C$ and $\log(C)$ would be considered ``confounders,'' though
$\log(C)$
would not be a confounder conditional on $C$, and likewise $C$ would
not be
a confounder conditional on $\log(C)$. We believe that this is in accord
with epidemiologic usage, though it would be peculiar to consider both $C$
and $\log(C)$ simultaneously, just as it would be peculiar to include
both $C$ and $\log(C)$ on a causal diagram. Second, if a variable $C$
is measured
with error, taking value $C^{\ast}$, and if the measurement error term
$\varepsilon=C^{\ast}-C$ were also represented on the causal diagram,
then, if $C$ were a confounder under Definition~\ref{def4}, $C^{\ast}$ and
$\varepsilon$ would
also both be confounders under Definition~\ref{def4}. We believe this is also in
accord with standard epidemiologic usage of ``confounder,'' though we
would in
practice rarely refer to $\varepsilon$ as a ``confounder'' since we
rarely have access to~$\varepsilon$. Once again, however, neither
$C^{\ast}$
nor $\varepsilon$ would be confounders conditional on~$C$. Finally,
suppose $C_{1}$ were height in meters and $C_{2}$ were weight in
kilograms and that $C_{1}$ and $C_{2}$ together sufficed to control for
confounding but neither
alone did; let $C_{3}=C_{1}/C_{1}^{2}$ be body mass index (BMI) and suppose
that controlling for $C_{3}$ alone sufficed to control for confounding. Then
under Definition~\ref{def4}, $C_{1}$, $C_{2}$ and $C_{3}$ would each be confounders,
though $C_{3}$ would not be a confounder conditional on $(C_{1},C_{2})$ and
likewise neither $C_{1}$ nor $C_{2}$ would be a confounder conditional
on $C_{3}$. Once again, we believe this is in accord with traditional
epidemiologic usage of ``confounder.''

Several implications hold between the different definitions of a
confounder as stated in the following result.

%pr7 #&#
\begin{proposition}\label{prop7}On a causal diagram, if a variable is a
confounder under Definition~\ref{def3}, then it is a confounder under
Definitions~\ref{def4},~\ref{def2} and~\ref{def1}; if under Definition~\ref{def4}, then under Definitions~\ref{def2}
and~\ref{def1}; if under Definition~\ref{def5}, then under Definitions~\ref{def6} and~\ref{def1}; if under
Definition~\ref{def6}, then under Definition~\ref{def1}. No other implications hold
without further assumptions.
\end{proposition}

\begin{pf} On a causal diagram, if a variable is a member of
every minimally sufficient adjustment set, it must be a member of a
minimally sufficient adjustment set (the existence of a minimally
sufficient adjustment set is guaranteed by the variables lying on a
causal diagram). Thus, if a variable is a confounder under Definition~\ref{def3}, then it is a confounder under Definition~\ref{def4}. Suppose a variable $C$
satisfies Definition~\ref{def4}, that is, is a member of some minimally
sufficient adjustment set $(X,C)$, but that it does not satisfy
Definition~\ref{def2}, that is, it is not on a backdoor path from $A$ to $Y$. By
Theorem~5 of Shpitser, VanderWeele and
Robins (\citeyear{ShpVanRob10}), $(X,C)$ blocks all backdoor paths
from $A$ to $Y$. If $C$ does not lie on a backdoor path from $A$ to
$Y$, then $X$ alone would block all backdoor paths from $A$ to $Y$,\vadjust{\goodbreak}
which would contradict that $(X,C)$ is a minimally sufficient
adjustment set. Thus, if $C$ is a confounder under Definition~\ref{def4}, it is
a confounder under Definition~\ref{def2}. That $C$ being a confounder under
Definition~\ref{def4} implies $C$ is a confounder under Definition~\ref{def1} follows
from the contrapositive of Corollary~4.1 of \citet{Rob97}. If $C$ is a
confounder under Definition~\ref{def5}, it must be a confounder under Definition~\ref{def6} because the only way $C$ can be a confounder under Definition~\ref{def5} is if
$ \sum_{x,c}\{E(Y|A=1,x,c)-E(Y|A=0,x,c)\}\operatorname{pr}(x,c)$ and $ \sum_{x}\{
E(Y|A=1,x)-E(Y|A=0,x)\}\operatorname{pr}(x)$ are not equal. If $C$ is not a confounder
under Definition~\ref{def1}, then for every $X$, $C$ is independent of $Y$ conditional on $(A,X)$ or of
$A$ conditional on $X$ and from this it easily follows that $ \sum_{x,c}\{E(Y|A=1,x,c)-E(Y|A=0,x,c)\}\operatorname{pr}(x,c)= \sum_{x}\{
E(Y|A=1,x)-E(Y|A=0,x)\}\operatorname{pr}(x)$ and thus that $C$ is not a confounder
under Definition~\ref{def6}. Thus, if $C$ is a confounder under Definition~\ref{def6}, it
must be a confounder under Definition~\ref{def1}.

We now argue that without further assumptions no other implications
between the definitions hold. The variable $C_{2}$ in Figure~\ref{fig4} could
satisfy Definition~\ref{def1} but does not satisfy Definition~\ref{def2}, so Definition~\ref{def1}
does not imply Definition~\ref{def2}. The variable $C_{3}$ in Figure~\ref{fig1} could
satisfy Definition~\ref{def1}, but does not satisfy Definitions~\ref{def3},~\ref{def4} or~\ref{def5}; thus,
Definition~\ref{def1} does not imply Definitions~\ref{def3},~\ref{def4} or~\ref{def5}. If $C$ is a
confounder under Definition~\ref{def1}, in general it will be under Definition~\ref{def6}
as well, but it may not because of cancellations due to scale-dependence.

If $C$ satisfies the conditions for Definition~\ref{def2} (i.e., lies on a
backdoor path from $A$ to $Y$), it will generally do so for Definitions
\ref{def1} and~\ref{def6} but may fail to do so because of failure or faithfulness or
cancellations due to scale-dependence. In the example given concerning
Property~\ref{prope2B} in Proposition~\ref{prop2}, the variable $C_{2}$ in Figure~\ref{fig2}
satisfied Definition~\ref{def2} but does not satisfy Definitions~\ref{def3},~\ref{def4} or~\ref{def5};
thus, Definition~\ref{def2} does not imply Definitions~\ref{def3},~\ref{def4} or~\ref{def5}.

It was shown above that if $C$ satisfies the conditions for Definition~\ref{def3}, it will satisfy the conditions for Definitions~\ref{def4},~\ref{def2} and~\ref{def1}. If $C$
satisfies the conditions for Definition~\ref{def3}, it will generally satisfy
the conditions for Definitions~\ref{def5} and~\ref{def6}, but it may not do so due to
scale-dependence.

It was shown above that if $C$ satisfies the conditions for Definition~\ref{def4}, it will satisfy the conditions for Definitions~\ref{def2} and~\ref{def1}. In Figure~\ref{fig3},
$C_{2}$ satisfies the conditions for Definition~\ref{def4} but not Definition~\ref{def3},
therefore, Definition~\ref{def4} does not imply Definition~\ref{def3}. If $C$ satisfies
the conditions for Definition~\ref{def4}, it will generally satisfy the
conditions for Definitions~\ref{def5} and~\ref{def6}, but it may not do so due to
scale-dependence.

It was shown above that if $C$ satisfies the conditions for Definition~\ref{def5}, it will satisfy the conditions for Definitions~\ref{def6} and~\ref{def1}. In the
example given concerning Property~\ref{prope2B} in Proposition~\ref{prop5}, the variable
$C_{2}$ in Figure~\ref{fig4} satisfied Definition~\ref{def5} but does not satisfy
Definitions~\ref{def2},~\ref{def3} or~\ref{def4}; thus, Definition~\ref{def5} does not imply Definitions~\ref{def2},
\ref{def3} or~\ref{def4}.

It was shown above that if $C$ satisfies the conditions for Definition~\ref{def6}, it will satisfy the conditions for Definition~\ref{def1}. The variable
$C_{2}$ in Figure~\ref{fig4} could satisfy Definition~\ref{def6} but does not satisfy
Definition~\ref{def2}, so Definition~\ref{def6} does not imply Definition~\ref{def2}. The variable
$C_{3}$ in Figure~\ref{fig1} could satisfy Definition~\ref{def6}, but does not satisfy
Definitions~\ref{def3},~\ref{def4} or~\ref{def5}; thus, Definition~\ref{def6} does not imply Definitions~\ref{def3},
\ref{def4} or~\ref{def5}.
\end{pf}

%f5 #&#
\begin{figure}

\includegraphics{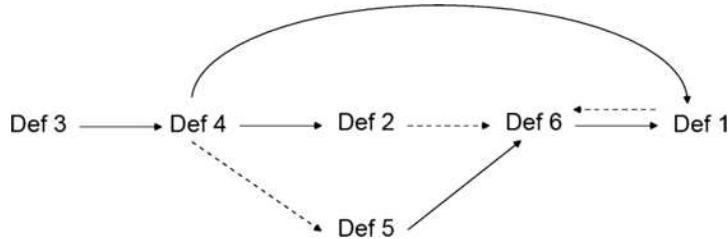}

\caption{Logical relationships
that hold among definitions. Dashed arrows indicate implications that
will
generally hold but may fail due to scale dependence of
definitions.}\label{fig5}
\end{figure}

The implications between the definitions are plotted in Figure~\ref{fig5}. Those
implications that will generally hold but may not hold because of
cancellations due to scale-dependence are indicated with dashed arrows.

The properties themselves that we have been considering also bear
certain relations to one another insofar as it is not difficult to show
that if Property~\ref{prope2A} is itself taken as the definition of a confounder,
then, on causal diagrams, this definition of a confounder also
satisfies Property~\ref{prope1}. This is because if $S$ denotes the set of all
nodes $C$ which obey Property~\ref{prope2A} and if $S$ is not a sufficient adjustment set
(so there is open backdoor path $\pi$ from $A$ to $Y$), then if we let
$W$ be all nondescendants of $A$ other than $A$ and noncolliders nodes
on $\pi$, if we choose a node $K$ on $\pi$ that does not contain
descendants of $A,$ then it is the case that $K$ satisfies Property~\ref{prope2A}, and is
not a part of $S$, which would be a contradiction.

Although it is the case that if Property~\ref{prope2A} is itself taken as the
definition of a confounder then this definition also satisfies Property~\ref{prope1} on causal diagrams, this does not hold generally within a
counterfactual framework. Note also that, even on causal diagrams, it
is not the case that Property~\ref{prope2A} implies Property~\ref{prope1}; a counterexample
to this was given in Proposition~\ref{prop3} for Definition~\ref{def3} which satisfies
Property~\ref{prope2A} but not Property~\ref{prope1}. Rather, if Property~\ref{prope2A} is itself taken
as the definition of a confounder, then, on causal diagrams, this
definition would satisfy Property~\ref{prope1} as well. This raises the question
as to whether Property~\ref{prope2A} itself could be taken as the definition of a
confounder, as such a definition would satisfy Property~\ref{prope2A} (by
definition) and Property~\ref{prope1} on causal diagrams. Although such a
definition would satisfy Properties~\ref{prope1} and~\ref{prope2A} on causal diagrams, it
would also follow from this definition that $C_{1}$ is\vadjust{\goodbreak} a confounder for
the effect of $A$ on $Y$ in Figure~\ref{fig1}, even though the effect $A$ on $Y$
is unconfounded without controlling for any covariates. This is because
if Property~\ref{prope2A} is taken as the definition of a confounder, then $C_{1}$
satisfies Property~\ref{prope2A} with $X$ taken as $C_{3}$. In general, however, if the
effect $A$ on $Y$ is unconfounded without controlling for any
covariates, we would probably simply say that there are no confounders
for the unconditional effect of $A$ on $Y$.

%s7 #&#
\section{Concluding remarks}\label{sec7}

The causal inference literature has provided a formal definition of
confounding with reference to distributions of counterfactual outcomes. The
literature now rightly emphasizes the concept of confounding control
over that of
a ``confounder.'' Nonetheless, the word ``confounder'' is
often still used among applied researchers and in this paper we have shown
that at least one formal counterfactual-based definition coheres with
the way in which the word is generally used.
We have considered a number of candidate proposals often arising
from more informal statements made in the literature. We have
considered whether
each of these definitions satisfies two properties, namely, (i) that on
any causal diagram, control for
all confounders so defined will control for confounding and (ii) any
variable qualifying as a
confounder under this criterion will in some context remove
confounding. Only one
of the definitions considered here satisfied both of these two properties.
We thus proposed that a pre-exposure covariate
$C$ be considered a confounder for the effect of $A$ on $Y$ if there
exists a
set of covariates $X$ such that the effect of the exposure on the
outcome is
unconfounded conditional on $(X,C)$ but for no proper subset of $(X,C)$ is
the effect of the exposure on the outcome unconfounded given the subset.
Equivalently, a confounder is a ``member of a minimally sufficient adjustment
set.'' This is closely related to the definitions concerning
confounders given in \citet{RobMor87}, though Robins and
Morgenstern suggest specifying the other variables for which control
might be made as well.
We have further provided a conditional analogue of the proposed
definition of a confounder; and we have proposed
that a variable that helps reduce bias but not eliminate bias be
referred to
as a ``surrogate confounder.'' The definition of a ``confounder'' above
is given rigorously in terms of counterfactuals and, we believe, is
also in accord with the intuitive
properties of a ``confounder'' implicitly presupposed by practicing
statisticians and
epidemiologists. From a more theoretical perspective, Definition~\ref{def4}, unlike
the other definitions, gives rise to elegant and useful results which itself
lends further support for its being taken as the definition of a confounder.

%sA #&#
\begin{appendix}\label{app}
\section*{Appendix}

\subsection*{Review of causal diagrams}

A directed graph consists of a set of nodes and directed edges among
nodes. A path is a sequence of distinct nodes connected by edges regardless
of arrowhead direction; a directed path is a path which follows the
edges in
the direction indicated by the\vadjust{\goodbreak} graph's arrows. A directed graph is acyclic
if there is no node with a sequence of directed edges back to itself. The
nodes with directed edges into a node $A$ are said to be the parents of $A$;
the nodes into which there are directed edges from $A$ are said to be the
children of $A$. We say that node $A$ is an ancestor of node $B$ if
there is a
directed path from $A$ to $B$; if $A$ is an ancestor of $B$, then $B$
is said
to be a descendant of $A$. If $X$ denotes a set of nodes, then $\operatorname{An}(X)$ will
denote the ancestors of $X$ and $\operatorname{Nd}(X)$ will denote the set of nondescendants
of $X$. For a given graph $G$, and a set of nodes $S$, the graph $G_{S}$
denotes a subgraph of $G$ containing only vertices of $G$ in $S$ and only
edges of $G$ between vertices in $S$. On the other hand, the graph
$G_{\overline{S}}$ denotes the graph obtained from $G $ by removing all edges
with arrowheads pointing to $S$. A node is said to be a collider for a
particular path if it is such that both the preceding and subsequent nodes
on the path have directed edges going into that node. A path between two
nodes, $A$ and $B$, is said to be blocked given some set of nodes $C$ if
either there is a variable in $C$ on the path that is not a collider
for the
path or if there is a collider on the path such that neither the collider
itself nor any of its descendants are in $C$. For disjoint sets of
nodes $A $, $B$ and $C$, we say that $A$ and $B$ are d-separated given
$C$ if every
path from any node in $A$ to any node in $B$ is blocked given $C$.
Directed acyclic graphs are sometimes used as statistical models to encode
independence relationships among variables represented by the nodes on the
graph [Lauritzen (\citeyear{La96})]. The variables corresponding to the nodes on a
graph are said to satisfy the global Markov property for the directed
acyclic graph (or to have a distribution compatible with the graph) if for
any disjoint sets of nodes $A,B,C$ we have that $A\perp\!\!\!\perp B|C$
whenever $A$ and $B$ are d-separated given $C$. The distribution of some
set of variables $V$ on the graph is said to be faithful to the graph if
for all disjoint sets $A,B,C$ of $V$ we have that $A\perp\!\!\!\perp B|C$
only when $A$ and $B$ are d-separated given $C$.

Directed acyclic graphs can be interpreted as representing causal
relationships. \citet{Pea95} defined a causal directed acyclic graph as a
directed acyclic graph with nodes $(X_{1},\ldots,X_{n})$ corresponding to
variables such that each variable $X_{i}$ is given by its nonparametric
structural equation $X_{i}=f_{i}(pa_{i},\varepsilon_{i})$, where
$pa_{i}$ are
the parents of $X_{i}$ on the graph and the $\varepsilon_{i}$ are mutually
independent. For a causal diagram, the nonparametric structural equations
encode counterfactual relationships among the variables represented on the
graph. The equations themselves represent one-step ahead counterfactuals
with other counterfactuals given by recursive substitution [see \citet{Pea09}
for further discussion]. A causal directed acyclic graph defined by
nonparametric structural equations satisfies the global Markov property as
stated above [\citet{Pea09}]. The requirement that the $\varepsilon_{i}$ be
mutually independent is essentially a requirement that there is no variable
absent from the graph which, if included on the graph, would be a
parent of
two or more variables [Pearl (\citeyear{Pea95}, \citeyear{Pea09})]. Throughout we assume the exposure
$A$ consists of a single node. A backdoor path from $A$ to $Y$ is a
path to
$Y$ which begins with an edge into $A$. A set of variables $X$ is said to
satisfy the backdoor path criterion with respect to $(A,Y)$ if no variable
in $X$ is a descendant of $A$ and if $X$ blocks all backdoor paths from $A$
to $Y$. \citet{Pea95} showed that if $X$ satisfies the backdoor path
criterion with respect to $(A,Y)$, then the effect of $A$ on $Y$ is
unconfounded given $X$, that is, $Y_{a}\perp\!\!\!\perp A|X$.

\subsection*{Empirical testing for confounders and confounding}

The absence of confounding conditional on a set of covariates $S$, that
is, $Y_{a}\perp\!\!\!\perp A|S$, is not a property that can be tested empirically
with data. One must rely on subject matter knowledge, which may sometimes
take the form of a causal diagram. Nonetheless, a~few things can be said
about empirical testing concerning confounding and confounders. For the sake
of completeness, we will consider each of Definitions~\ref{def1}--\ref{def6}. It is
possible to
verify empirically whether a variable is a confounder under Definition~\ref{def1}
since the definition refers to observed associations; however, it is not
possible, without further knowledge, to empirically verify that a variable
does not satisfy Definition~\ref{def1} because a variable may satisfy Definition~\ref{def1}
for some $X$ that involves an unmeasured variable $U$. One would have to
know that data were available for all variables on a causal diagram to
empirically verify that a variable was a nonconfounder under Definition~\ref{def1}.
Because of this, even though Definition~\ref{def1} satisfies Property~\ref{prope1} under
faithfulness, this cannot
be used as an empirical test for confounding since (i) we cannot empirically
verify that a variable is a nonconfounder under Definition~\ref{def1} and (ii) we
cannot empirically verify whether faithfulness holds.

Without further assumptions, we cannot empirically verify that a
variable is
a confounder or a nonconfounder under Definition~\ref{def2} because Definition~\ref{def2}
makes reference to backdoor paths. Whether a variable lies on a backdoor
path cannot be tested empirically without further assumptions; one would
have to know the structure of the underlying causal diagram. Likewise, for
Definitions~\ref{def3} and~\ref{def4}, one would need to know all minimally sufficient
adjustment sets, which itself would require checking the ``no confounding''
condition $Y_{a}\perp\!\!\!\perp A|S$, which is, as noted above, not
empirically testable; though see below for some qualifications. For
Definition~\ref{def5}, we could empirically reject the inequality in Definition~\ref{def5} for
observed $X$ if $\sum_{x,c}\{E(Y|A=1,x,c)-E(Y|A=0,x,c)\}\operatorname{pr}(x,c)=\sum_{x}\{E(Y|A=1,x)-E(Y|A=0,x)\}\operatorname{pr}(x)$. However, we cannot
empirically reject the inequality in Definition~\ref{def5} for unobserved $X$
and we,
moreover, cannot empirically verify the inequality in Definition~\ref{def5}
because $E(Y_{1})-E(Y_{0})$ will not in general be empirically
identified if there
are unobserved variables. We can verify empirically whether a variable
is a confounder under Definition~\ref{def6}
since the definition refers to only observed variables; however, it is not
possible, without further knowledge, to empirically verify that a variable
does not satisfy Definition~\ref{def6} because a variable may satisfy Definition~\ref{def6}
for some $X$ that involves an unmeasured variable~$U$. One would have to
know that data were available for all variables on a causal diagram to
empirically verify that a variable was a nonconfounder under Definition~\ref{def6}.
Because of this we cannot empirically
verify that a variable is a nonconfounder under Definition~\ref{def6}.

Determining whether a variable is a confounder requires making untestable
assumptions. The only real progress that can be made with empirical testing
for confounders is by making other untestable assumptions that logically
imply a test for assumptions we care about. For example, suppose we assume
we have some set $S$ that we are sure constitutes a sufficient adjustment
set. In this case, we can sometimes remove variables as unnecessary for
confounding control. In particular, \citet{Rob97} showed that if we knew
that for covariate sets $S_{1}$ and $S_{2}$ we had that $Y_{a}\perp
\!\!\!\perp A|(S_{1},S_{2})$, then we would also have that $Y_{a}\perp
\!\!\!\perp A|S_{1}$ if $S_{2}$ can be decomposed into two disjoint
subsets $T_{1}$ and $T_{2}$ such that $A\perp\!\!\!\perp T_{1}|S_{1}$
and $Y\perp
\!\!\!\perp T_{2}|A,S_{1},T_{1}$. Both of these latter conditions are
empirically testable. \citet{Genetal01} provide some analogous results for
the effect of exposure on the exposed. \citet{VanShp11} note
that if for covariate set $S$ we have that $Y_{a}\perp\!\!\!\perp A|S$, then
if a backward selection procedure is applied to $S$ such that variables are
iteratively discarded that are independent of $Y$ conditional on both
exposure $A$ and the members of $S$ that have not yet been discarded, then
the resulting set of covariates will suffice for confounding control. They
also show that under an additional assumption of faithfulness, if, for
covariate set $S$, we have that $Y_{a}\perp\!\!\!\perp A|S$, then if a
forward selection procedure is applied to $S$ such that, starting with the
empty set, variables are iteratively added which are associated with $Y$
conditional on both exposure $A$ and the variables that have already been
added, then the resulting set of covariates will suffice for confounding
control. Note, however, all of these results require knowledge that for some
set $S$, $Y_{a}\perp\!\!\!\perp A|S$, which is not itself empirically
testable without experimental interventions.

\end{appendix}

\section*{Acknowledgments}
The authors thank Sander Greenland,
James Robins and Miguel Hern\'{a}n for helpful comments on this
paper.

% imsref loaded by akundreckaite, 2012-11-21 15:10:35

%suskaldyti doi

\printaddresses

\end{document}